\documentclass[12pt]{iopart}

\usepackage[
    backend=biber,
    style= authoryear,
    maxbibnames=2,
    url=false,
    sorting=nyt,
    natbib=true,
    dashed = false
  ]{biblatex}

\DefineBibliographyStrings{english}{
    andothers = {\etal}
}

\usepackage[ colorlinks=true,
    citecolor=blue,
    linkcolor = black,
    urlcolor=blue]{hyperref}

\usepackage[utf8]{inputenc} \usepackage[english]{babel} \usepackage[T1]{fontenc}  
\usepackage{lmodern} 
\usepackage{graphicx} 
\usepackage{url} 
\usepackage{calc}
\usepackage{multimedia} \usepackage{amsmath,amsfonts,amssymb}
\usepackage{pifont} 
\usepackage{textcomp} 

\usepackage{array}

\begin{document}

\title[A TOF-Based Reconstruction for Real-Time Prompt-Gamma Imaging]{A Time-Of-Flight-Based Reconstruction for Real-Time Prompt-Gamma Imaging in Protontherapy}
\author{Maxime Jacquet$^{1,6}$, Sara Marcatili$^{1,6}$,  Marie-Laure Gallin-Martel$^1$, Jean-Luc Bouly$^1$, Yannick Boursier$^3$, Denis Dauvergne$^1$, Mathieu Dupont$^3$, Laurent Gallin-Martel$^1$, Jo\"{e}l H\'{e}rault$^2$, Jean Michel Létang$^4$, Daniel Manéval$^2$, Christian Morel$^3$, Jean-François Muraz$^1$, \'{E}tienne Testa$^5$}
\address{
$^1$ Universit\'{e} Grenoble Alpes, CNRS, Grenoble INP, LPSC-IN2P3 UMR 5821, 38000 Grenoble, France\\
$^2$ Centre Antoine Lacassagne, 06200 Nice, France\\
$^3$ Aix-Marseille Univ, CNRS/IN2P3, CPPM, Marseille, France\\
$^4$ University of Lyon, INSA-Lyon, Universit\'{e} Claude Bernard Lyon 1, UJM-Saint Etienne, CNRS, Inserm, CREATIS UMR 5220, U1206, F-69373 Lyon, France\\
$^5$  Univ.  Lyon, Univ. Claude Bernard Lyon 1, CNRS/IN2P3, IP2I Lyon, F-69622, Villeurbanne, France \\
$^6$ Authors to whom any correspondence should be adressed.
}
\ead{marcatili@lpsc.in2p3.fr, majacquet@lpsc.in2p3.fr}

\begin{abstract} 
We are currently conceiving,  through (MC) simulation, a multi-channel gamma detector array (TIARA for Time-of-flight Imaging ARrAy) for the online monitoring of protontherapy treatments. By  measuring the Time-Of-Flight (TOF) between a beam monitor placed upstream and the Prompt-Gamma (PG) detector, our
 goal is to reconstruct the PG vertex distribution to detect a possible deviation of proton beam delivery. In this paper, two non-iterative reconstruction strategies are proposed. The first  is based on the resolution of an analytical formula describing the PG vertex distribution in 3D (PG vertex reconstruction). Here, it was resolved under a one-dimensional approximation in order to measure a potential proton range shift along the beam direction. The second is based on the calculation of the Centre-Of-Gravity (COG) of the TIARA pixel detectors counts and also provides 3D information on a possible beam displacement (COG method). \\
The PG vertex reconstruction was evaluated in two different scenarios. A  coincidence time resolution  of 100 ps (rms) can be attained in single proton regime (operating a reduction of the beam current) and  using an external beam monitor to provide a start trigger for the TOF measurement. 
Under these conditions, MC simulations have shown that a millimetric proton range shift sensitivity can be achieved at 2$\sigma$ with 10$^{8}$ incident protons. This level of accuracy would allow to act in real-time if the treatment does not conform to treatment plan.
 A worst case scenario of a 1 ns (rms) TOF resolution was also considered to demonstrate that a degraded timing information can be compensated by increasing the acquisition statistics: in this case, a 2 mm range shift  would be detectable at 
 2$\sigma$ with  10$^{9}$ incident protons.\\
 At the same time, the COG method has shown  excellent capabilities of detecting lateral beam displacements: a 2 mm sensitivity was found at  2$\sigma$ with  10$^{8}$ incident protons.
\end{abstract}
\newpage
\section{Introduction}
Hadrontherapy is a non-invasive radiotherapy modality using charged ions to selectively irradiate tumours.
With more than one hundred clinical centres operating worldwide in 2020, this technique has known a remarkable development in recent years (Dosanjh \etal 2018). 
Compared to conventional irradiation methods (X-ray and electron beams),  hadrontherapy presents a  distinctive depth dose deposition profile, with a maximum  at the end of the ion range (Bragg peak). 
This feature provides, in principle, a very high ballistic precision and tumour coverage together with a limited irradiation of healthy tissues located in both proximal and distal regions of the field.
However, in the clinical practice, the potential accuracy of hadrontherapy is compromised by different sources of uncertainties such as transient modifications of the anatomy (tumour mass reduction, weight
loss, daily changes of internal cavity filling), errors in the determination of  patient tissue composition, physiological movements of organs, or simply patient mispositioning (Paganetti 2012). 
As a consequence, safety margins are routinely applied to the tumour volume at treatment planning: in case of
deep-seated tumour, these  margins can be as large as 1 cm, thus limiting treatment selectivity (Paganetti 2012). For the same reasons,  irradiation fields presenting an  organ at risk in the tumour distal region are usually avoided in favor of multiple irradiation fields, with the consequence of increasing the dose to healthy tissues in the proximal region (Knopf and Lomax 2013).\\
The use of a detection system to measure the hadron range in real-time is today recognised as essential for the improvement of hadrontherapy efficacy  (Pausch \etal 2020). Over the last decade, many research groups have developed a large variety of monitoring systems, which take advantage from the existing correlation between the proton range and the physical properties of secondary particles produced by nuclear interactions within the patient (Kraan 2015, Krimmer \etal 2018). 
In the case of protontherapy, these include in beam or post-treatment Positron Emission Tomography (PET) exploiting 511 keV gamma rays from $\beta^{+}$ emitting fragments  (Enghardt \etal 2004, Bisogni \etal 2016, Ferrero \etal 2018)
and a large variety of systems based on the detection of Prompt-Gamma (PG) rays (Krimmer \etal 2018, Parodi \etal 2018).  
Detectors falling in the last category focus on the measurement of different physical variables to infer information on the beam path
in the distal and/or transverse directions.\\ 
In PG Imaging (PGI), the longitudinal distribution of PG emission vertices  is directly measured using collimated gamma detectors (Smeets \etal 2012, Xie \etal 2017, Perali \etal 2014, Priegnitz \etal 2015) while the 3D PG vertex distribution  can be accessed using Compton cameras (Roellinghoff \etal 2011, Kishimoto \etal 2015, Krimmer \etal 2015a, Thirolf \etal 2017, Muñoz \etal 2017, Draeger \etal 2018). A Time-Of-Flight (TOF) information can be added to these systems in order to reject time-uncorrelated
particles (mainly neutrons) and therefore to improve the
sensitivity of the proton range measurement (Krimmer \etal 2015b, Pinto \etal 2014).
To the best of our knowledge, those based on collimated cameras are the sole PGI approaches capable of providing real-time monitoring of the proton range (Richter \etal 2016, Xie \etal 2017): the most promising prototype allows detecting range shifts below 4 mm at 2$\sigma$ with 0.5$\times$10$^{8}$ incident protons at 100 MeV (Perali \etal 2014). 
\\
In PG Spectroscopy (PGS), the energy and target material dependencies of nuclear cross sections are taken into account
in the analysis of PG energy spectra in order to obtain information on the residual proton range (with respect to the centre of the gamma detector Field-Of-View (FOV)) and the target chemical composition (Verburg and Seco 2014). A detection system based on this principle was recently tested on phantoms under clinically relevant irradiation conditions: the absolute proton range was measured with millimetric precision at 2$\sigma$ with a statistics of $\sim10^{9}$ incident protons (Hueso-González \etal 2018).\\
Another promising technique is the PG Peak Integral (PGPI) approach proposed by Krimmer \etal (2017). It consists in comparing the ratio of integral counts from multiple gamma detectors arranged around the target in order to detect a possible 3D beam displacement. A TOF information  is also included to discriminate PGs produced in the target from those produced outside. According to Monte Carlo (MC) simulations, a few per cent of variation in the registered PG rate can be detected with 10$^8$ incident protons, corresponding to a proton-range shift of a few millimeters.\\
In this work, we focus on the detection of proton range shifts through the measurement of the TOF of PGs. The so-called Prompt-Gamma Timing (PGT) approach was first proposed by Golnik \etal (2014) and exploits the existing correlation between ion ranges and TOF distribution characteristics. In Hueso-González \etal (2015), the authors showed how the proton range can be estimated from the momenta of the PG TOF spectrum acquired by a fast gamma detector (e.g BaF$_{2}$ or LaBr$_{3}$ crystal) placed at backward detection angles relative to the beam direction.  The precision of this technique directly depends on the time resolution of the detection system, and since the TOF is measured with respect to the beam RF signal, for cyclotrons the time resolution is ultimately limited by the time-width of the proton bunch  (Petzoldt \etal 2016). According to Pausch \etal (2016), in a worst case scenario of 2 ns (FWHM) time resolution, a 5 mm shift sensitivity would be achievable from a PG TOF spectrum including 10$^{4}$ events.
More recently, the CLaRyS collaboration has  proposed the use of a fast beam-tagging hodoscope (Gallin-Martel \etal 2018, Marcatili \etal 2020)  with the aim to provide a precise start time for the TOF measurement and therefore improve the sensitivity of the PGT  technique. Besides, the use of a beam monitor would allow detecting any possible beam instability in real-time.\\

In this paper we propose a novel approach to real-time monitoring, based on TOF-resolved PGI, namely PGTI as PG Time Imaging.  We are currently conceiving, through MC simulation, a detection system (TIARA for Time-of-flight Imaging ARrAy) to measure the 3D distribution of PG vertices in real-time by exclusively measuring the PG TOF.
TIARA will be read out in time coincidence with a fast beam monitor placed upstream the patient, with a targeted Coincidence Time Resolution (CTR) of 100 ps (rms). This CTR may be achieved reducing the beam intensity at the  beginning of the irradiation (single proton regime) as proposed by Dauvergne \etal (2020) and Marcatili \etal (2020), in order to tag in time each single proton and thus overcome the limitation that an extended bunch time-width imposes
 to the system CTR. Our goal is to verify that the treatment is consistent with the treatment plan already within the first (or first few) irradiation spot(s) by comparing the PG vertex distribution measured by TIARA to the one calculated under Treatment Planning System (TPS) conditions. \\
The development of a dedicated, fully-3D, reconstruction algorithm is currently on-going. This paper proposes two different simplified reconstruction strategies.  The first one is based on the solution of an analytical reconstruction formula describing the PG vertex distribution in the target, and it is discussed in section \ref{AR_section}. The second, based on the calculation of the center of gravity of TIARA pixel detectors, is described in section \ref{COG_sec}. 
\begin{figure}[!h] 
\centering 
\includegraphics[width = 0.5\textwidth]{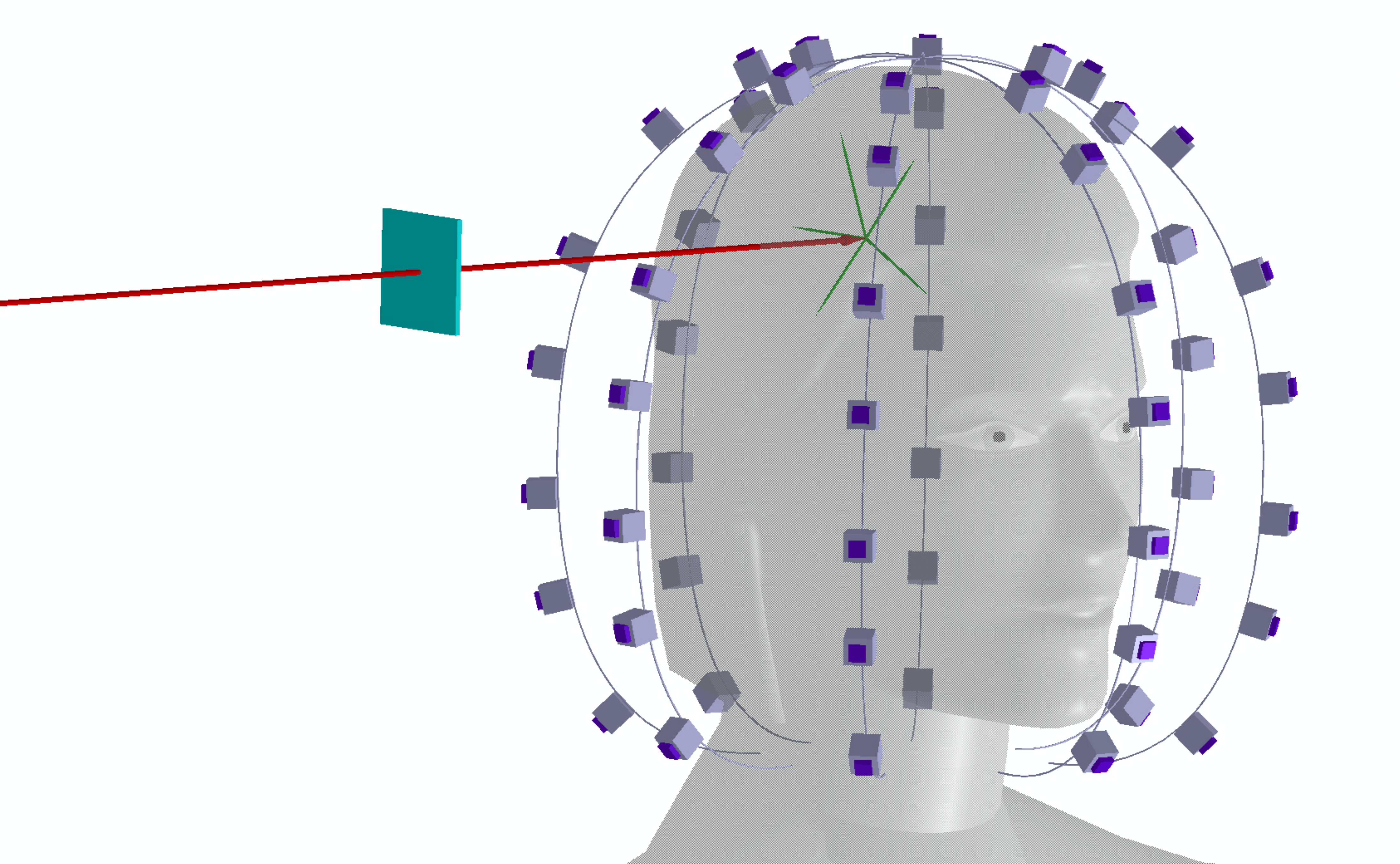} 
\caption{Schematic view of a possible TIARA design: the pixels are placed around the target at fixed positions. Pixels coordinates can be established patient-wise at treatment planning in order to clear one or more entrance windows for the proton beam (red line). TIARA will be read in time coincidence with a beam monitor (light blue). \label{tiara_design}}.
\end{figure}
\section{Material and methods}
\subsection{TIARA: a Time-of-flight Imaging ARrAy}
\begin{figure}[!h] 
\centering 
\includegraphics[width = 1\textwidth]{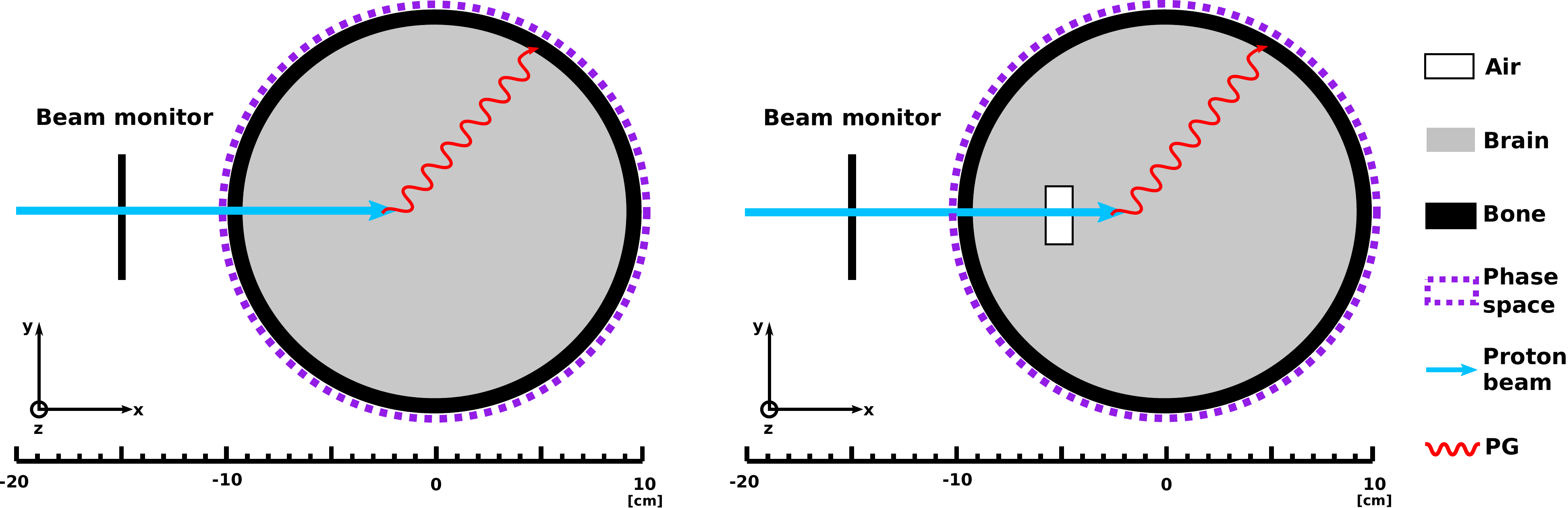} 
\caption{Left:  geometry  simulated with Geant4 to validate the analytical reconstruction algorithm. The impinging 100 MeV proton beam  is represented in blue. In red, a PG created along the beam path and recorded in the phase space is also shown. A diamond based beam monitor (represented in black) placed upstream of the target was considered. Right: the same geometry is displayed, with the insertion of an air cavity of variable size and fixed position at (-5.5,0,0).} \label{tiara_simulation}
\end{figure}

TIARA will be composed by approximately 30 independent gamma detectors (pixels) providing a 3D coverage around the patient. The number of pixels and their spatial arrangement can be optimised according to patient anatomy. A schematic view of a possible TIARA design is shown in figure~\ref{tiara_design}. In this work, a spherical head geometry (cf figure~\ref{tiara_simulation},~left) with a uniform angular coverage is considered.
Each pixel will consist of  a $\sim$ 1 cm$^{3}$ Cherenkov radiator (i.e. PbF$_{2}$) read out by a Silicon PhotoMultiplier (SiPM) (Marcatili \etal 2019). The objective is to exploit the Cherenkov light emitted by high-energy secondary electrons  (resulting from the PG interactions in the detector) propagating in a highly-refractive medium.
The advantage of using Cherenkov radiators is twofold: first, the Cherenkov process is inherently much faster than scintillation, allowing the improvement of the TOF resolution; second, many Cherenkov radiators present higher attenuation coefficient than commonly used scintillators, resulting in a higher detection efficiency for the same detection volume. \\
With the pixel spatial coordinates  defined at the beginning of the treatment, each pixel will not only provide the PG time of arrival, but also the PG hit coordinates with a spatial resolution determined by the spatial extension of the pixel itself.  A very poor energy resolution is expected for 1 cm$^{3}$ Cherenkov radiators whose limited size prevents the full absorption of PGs.
 However, it was already shown, exploiting the fast signal component of a 38 cm$^{3}$ BaF$_{2}$ scintillator (not suitable for spectroscopy), that an accurate determination of the PG energy is not essential to detect a proton range shift of a few millimeter when a high time resolution PGT monitoring approach is implemented (Marcatili \etal 2020).\\ 
The proton plus PG TOF will be obtained exploiting the time coincidence with a beam monitor placed upstream the target (see figure~\ref{tiara_design}). A CTR of 100 ps (rms) was already obtained in a previous experiment (Marcatili \etal 2020) employing the aforementioned BaF$_{2}$ gamma detector (presenting a detection efficiency close to the one targeted for TIARA), acquired in time coincidence with a single-crystal diamond detector acting as beam monitor. 
\subsection{PG vertex reconstruction}\label{AR_section}
The TOF measured by TIARA for each event can be written as the sum of three different components: the proton TOF (\textit{T}$_{proton}$) between the beam monitor and the nuclear interaction vertex; the target nucleus de-excitation time (\textit{T}$_{decay}$) associated to the emission of a PG; and the TOF between the interaction vertex and the PG detection point (\textit{T}$_{PG}$).\\
\begin{equation} 
TOF = T_{proton}(\textbf{r}_{v}) + T_{decay} + T_{PG}(\textbf{r}_{d}, \textbf{r}_{v}) 
\label{eq_TIARA} 
\end{equation} 
where $\textbf{r}_{d}$ (\textit{x}$_d$, \textit{y}$_d$, \textit{z}$_d$) is the PG hit coordinate vector  in the gamma detector and $\textbf{r}_{v}$  (\textit{x}$_v$, \textit{y}$_v$, \textit{z}$_v$) is the coordinate vector of the PG vertex.
Since nuclear de-excitation half-lives are below $\sim$ 1 ps for most biologically relevant nuclei (Kozlovsky \etal 2002), the contribution of $T_{decay}$ in equation~\ref{eq_TIARA} can be safely neglected at the time-scale of the fastest gamma-ray detector. \\
The gamma-ray TOF (T$_{PG}$) can be analytically determined from the geometrical distance between the PG vertex and the PG hit position:
\begin{equation}
T_{PG} = \frac{1}{c}  \|\textbf{r}_d -\textbf{r}_v\| ~~.
\end{equation}
From these considerations, equation~\ref{eq_TIARA} becomes: 
\begin{equation} 
\small TOF =
T_{proton}(\textbf{r}_{v})  + \frac{1}{c}  \|\textbf{r}_d -\textbf{r}_v\|
\label{eq_TIARA_tmp_2} 
\end{equation}
describing the PG vertex distribution in the three-dimensions.
With the further assumption that the proton beam lays on the \textit{x}-axis,
$y_{v}$ and $z_{v}$ can be set to zero and $T_{proton}$ only becomes a function of $x_{v}$. Equation~\ref{eq_TIARA_tmp_2} is therefore reduced to one dimension to allow the calculation of the PG emission point along the beam path:
\begin{equation} 
   TOF - \frac{\sqrt{(x_d -x_v)^2 + y_d^2 + z_d^2}}{c} - T_{proton}(x_{v}) = 0.
   \label{eqfinale} 
\end{equation}
Once $T_{proton}$ is determined, equation~\ref{eqfinale} can be resolved through a binary search of zeros to obtain the distal PG vertex coordinate $x_v$. 
\subsubsection{Determination of  $T_{proton}(x_{v})$.}\label{Tproton}  
The proton arrival time  cannot be  experimentally determined on an event-by-event basis, therefore a MC simulation based on the Geant4 toolkit (10.4.patch02 release) (Agostinelli \etal 2003) and the predefined QGSP\_BIC\_HP\_EMY physics list\footnote{https://geant4.web.cern.ch/node/155} was developed to calculate the $T_{proton}(x_{v})$ term in equation~\ref{eqfinale}. The simulated geometry consists in a 10 cm radius  spherical head  composed of homogeneous brain tissue surrounded by a 0.7 cm thick skull as shown in figure~\ref{tiara_simulation},~left, and surrounded by air. The origin of the coordinate system is the phantom centre.
A 100 MeV proton beam originating from a point-like source placed at ($-$20, 0, 0) cm impinges on the head along the \textit{x}-axis; the resulting range is 6.9 cm. A 50$\times$50$\times$0.3 mm$^3$ diamond-based beam monitor was placed upstream at ($-$15, 0, 0) cm. 
Protons' arrival time was scored every 0.1 mm along the beam axis on an event-by-event basis in order to compute the average of $T_{proton}$ as a function of beam penetration in the target and the associated statistical uncertainties. 
\subsubsection{PG vertex reconstruction validation.}
The geometry described in figure~\ref{tiara_simulation},~left was also used to validate the proposed method (equation~\ref{eqfinale}) for the reconstruction of the PG vertex profile. 4$\times$10$^{8}$ 100 MeV protons were simulated. A spherical detection surface was implemented as a 4$\pi$ phase space surrounding the phantom (represented by a dashed line in figure~\ref{tiara_simulation},~left) to record the  energy, position, time of arrival, vertex coordinate and ID of each impacting particle. 
The use of a phase space allowed evaluating the potential of the reconstruction algorithm without considering contingent detector characteristics as pixel size, material and geometrical arrangement, that have yet to be precisely defined (TIARA pixel R\&D is currently on-going). \\
For validation purposes, only unscattered PGs generated from primary protons were considered, as they are the only ones whose  TOF and  vertex coordinates are actually  correlated. A selection of the incident particle energy above 2 MeV  was also applied: this cut allows rejecting 511 keV emissions and the extremely delayed gamma-ray line from the 0.718 MeV $^{10}$B de-excitation (Kozlovsky \etal 2002).
The physical quantities scored in the phase space were used to solve equation~\ref{eqfinale} and reconstruct a longitudinal PG vertex profile. This profile was compared to the distribution of the PG $x$ vertex coordinates directly issued from the simulation and considered as the ground truth.
\subsubsection{Detector modeling.} \label{det_resp}
\begin{figure}[!h] 
\centering 
\includegraphics[width=1.\textwidth]{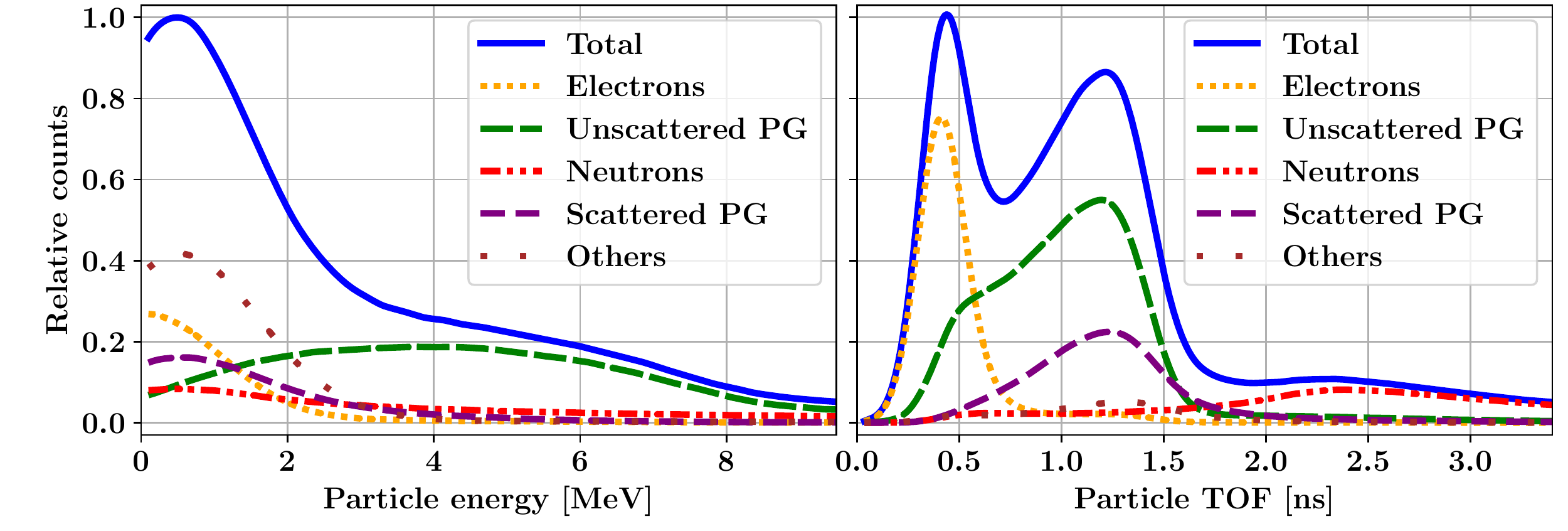} %
\caption{Energy  (left)
and TOF  (right) spectra of each particle type scored in the phase space. "Others" emissions mainly include 511 keV gamma-rays, but also PGs from secondary neutrons and protons. } 
\label{selection_criteria} 
\end{figure}

The MC simulations carried out for validation purposes considered an ideal detection system with perfect time, energy and spatial resolutions.
In order to mimic the expected response of TIARA, simulated data were first convolved on an event-by-event basis with realistic time and energy resolutions. Considering that PGs are not supposed to be fully absorbed in one TIARA pixel because of its limited size, a very poor energy resolution of 1 MeV (rms) was arbitrarily assumed. 
 Two different time resolutions were explored: the targeted CTR of 100 ps (rms) that would only be achievable in single proton regime; and a value of 1 ns (rms) corresponding to a CTR realistically achievable with a C-230 cyclotron at nominal beam intensity (Petzoldt \etal 2016, Werner \etal 2019).
Energy (left) and TOF (right) distributions, respectively corresponding to energy and time resolutions of 1 MeV and 100 ps, are shown in figure~\ref{selection_criteria} for the various secondary particles produced in the target. These distributions can be used to establish data selection criteria for optimal background rejection.
In the  energy distribution (left), contributions of scattered and unscattered PGs from primary protons are presented separately. The scattered PG (purple) spectrum presents a lower mean energy than the spectrum of unscattered PGs (green): an energy selection above 2 MeV (\textit{Ecut}) could therefore allow rejecting most of scattered PGs whose TOF is not correlated to the vertex coordinate. The same energy cut would also allow rejecting the fast electrons generated by the beam-induced ionisation in air.
At the same time, the TOF distribution (right) shows that a TOF selection below 1.6 ns (\textit{Tcut}) would allow rejecting the neutron contribution. While the \textit{Ecut} was systematically applied to the data analysis described in this paper, the \textit{Tcut} was not always necessary as the neutron signal   lays outside the PG time-spectrum (in green) leading to a natural neutron time-discrimination.\\
Since the detector geometry is not taken into account in the MC simulation, an off-line data selection strategy was adopted to reproduce a realistic detection efficiency for TIARA. In order to take into account the TIARA angular coverage, the phase space was reduced to thirty 1 cm$^{2}$ detection areas (pixels) homogeneously distributed  over the sphere (leaving an opening at beam entrance); only simulated particles hitting these areas were kept for further analysis. With this procedure, a geometrical efficiency of  $\sim$ 2.4$\%$ was obtained. At the same time, all particle hit positions falling within the pixel area were forced to the pixel center coordinates. %
For each pixel, a constant, conservative detection efficiency of 26.6$\%$ was set, which roughly corresponds to the interaction probability of a 4 MeV PG within a 1 cm thick PbF$_{2}$ crystal; at lower and higher energies, the photon interaction probability in  PbF$_{2}$ is higher respectively because of the increased Compton and pair production cross sections. With this additional hypothesis, simulated data can be selected to reproduce the response of a system with an overall detection efficiency of 0.6$\%$. \\ 
\subsubsection{Sensitivity of the PG vertex reconstruction.}\label{cavity}
The capability of the simplified 1D analytical reconstruction algorithm to detect a proton range shift along the beam direction was  investigated through MC simulation. The goal was to observe a possible discrepancy of the proton range during the irradiation with respect to the treatment planning data (reference irradiation conditions), for different numbers of incident protons and two different CTRs (100 ps and 1 ns rms). \\ \\ 
\textit{Simulation of reference irradiation and range deviations}\\
 In a prospective clinical application, TIARA response can be simulated before treatment considering the patient scans used for treatment plan, and the planned irradiation fields, to obtain a PG vertex distribution in reference conditions.
In this study, the geometry described in section 2.1 was modified to define the reference simulation: a 1$\times$2$\times$2 cm$^3$ box-shaped air cavity  centred at ($-$5.5,0,0) cm (i.e. 2.4 cm before the Bragg peak for 100 MeV impinging protons) was included in the head, as displayed in figure~\ref{tiara_simulation},~right. First, 4$\times$10$^{8}$  primary protons originating from a point-like beam laying on the \textit{X}-axis,  were simulated  to obtain the specific $T_{proton}(x_{v})$ curve for this geometry  following the approach described in section \ref{Tproton}. Then, secondary particles' time, energy and hit position scored in the phase space were used to reconstruct the PG vertex profile according to equation~\ref{eqfinale}, after applying the data selection described in section \ref{det_resp}. After a detection efficiency of 26.6\% is taken into account, the  PG profile simulated with a perfect detector and 4$\times$10$^{8}$  incident protons can also be interpreted as the profile achievable with a realistic detector and 1.5 $\times$10$^{9}$ incident protons (\textit{TPS statistics}).
Two different system CTRs were considered (100 ps and 1 ns rms) to analyse both simulated data from the phase space and the $T_{proton}(x_{v})$ curve. \\
In order to reproduce a distal proton range shift at treatment time, the air cavity thickness was progressively increased to 1.1, 1.2, 1.3, 1.4 and 1.5 cm with the cavity centre fixed at ($-$5.5, 0, 0) cm: considering the low proton energy loss in air, 1 mm of air roughly induces 1 mm of range shift. 
Data resulting from these five simulations with variable air cavity thickness were analysed considering  two  time resolutions and different numbers of incident protons (\textit{monitoring statistics}). In the case of a 100 ps CTR, data selections corresponding to 10$^{7}$ and 10$^{8}$ incident protons were performed, as this level of time resolution would only be achievable if a beam intensity reduction is applied at the beginning of the treatment for a very limited time. Since a 1 ns (rms) CTR may realistically be achieved at nominal beam intensity with a C-230 cyclotron, without reducing the beam current,  a higher \textit{monitoring statistics} of 10$^{9}$ protons was explored in this case.
For all scenarios, the T$_{proton}$ curve calculated under TPS conditions
was convolved with the corresponding time resolution and used for the reconstruction of PG vertex profiles for different air cavity thicknesses. \\\\
\textit{PG fall-off retrieval}\\
The difference in proton range obtained from the reference (TPS conditions) and the treatment simulations was measured using the method already described in Marcatili \etal (2020) and here summarised.
In the reconstructed PG profile  at \textit{TPS statistics} for the 1 cm air cavity, the reference x value ($x_{TPS}$) corresponding to the distal maximum of the distribution (which is ultimately correlated to the Bragg peak position) was defined: this value, can always be defined before treatment, independently of any possible variation in the PG profile shape occurring at treatment time.
At the same time, the profiles corresponding to the six different air cavity thicknesses (from 1.0 to 1.5 cm) were integrated to minimise the statistical fluctuations: the TPS integrated profile at \textit{TPS statistics} is defined as $f_{TPS}(x)$, while the integrated profiles for the $i^{th}$ treatment at \textit{monitoring statistics} is defined as $f_{i}(x)$, with $i$ representing the induced proton shift (from 0 to 5 mm). 
A reference value defined as  $y_{TPS}$ = $f_{TPS}$($x_{TPS}$) is then calculated on the TPS integrated profile.
For each air cavity thickness, the proton range shift is  measured as the $x$-axis distance ($d_{i}$) between the $x_{TPS}$  and $f_{i}^{-1}(y_{TPS})$.\\\\
\textit{Determination of the PG vertex reconstruction sensitivity} \label{sensitivity}\\
In order to estimate TIARA sensitivity to a proton range shift, $5\times10^4$ experiments were simulated in TPS conditions (1 cm air cavity), and for the five modified cavity thicknesses (from 1.1 to 1.5 cm in 1 mm steps).  For each experiment, a \textit{monitoring statistics} of $10^7$, $10^8$ (100 ps rms CTR hypothesis) or $10^9$ (1 ns rms CTR hypothesis) incident protons was randomly selected from the simulation dataset corresponding to $1.5\times10^9$ protons. Even though they are not unconnected, after event-by-event convolutions with the system time and energy resolutions, these sub-samples can be safely assumed to be statistically independent. Successively, the distance $d_i$ was calculated for each simulated experiment carried out in treatment conditions ($0\leq i \leq5$) to build the corresponding probability density functions (pdf).\\
With this procedure, six different normalised pdfs were obtained, one for each induced proton range shift (from 1 to 5 mm), plus the one corresponding to a treatment without anatomical variation with respect to treatment planning: the latter is considered as the reference pdf and defines the H0 hypothesis of no shift. The standard deviation of these distributions has been interpreted as the statistical errors on the range shift measurement.\\
Once the Confidence Level (CL), and therefore the type I error ($\alpha$) is fixed on the H0 pdf, the corresponding type II error ($\beta$) is calculated for the others as described in Marcatili \etal (2020): $\beta$ corresponds to the probability of not detecting a shift when a shift is actually present. %
Thus, a given proton range shift is considered detectable if, for a type I error ($\alpha$) fixed, the corresponding type II error ($\beta$) is lower than $\alpha$. 
In this work, the analysis was carried out for  two different CLs: 1$\sigma$ and 2$\sigma$.
%
\subsection{Centre of gravity reconstruction}\label{COG_sec}
A complementary reconstruction approach based on the calculation of the Centre-Of-Gravity (COG) of TIARA pixels coordinates weighted by the counts acquired in each pixel was also explored.
The COG coordinate vector $\textbf{r}_{COG}$ is calculated as:
\begin{equation} 
\textbf{r}_{COG} = \frac{1}{N}  \sum\limits_{i=1}^{N_{Det}} \textbf{r}_{i}  n_{i}
\label{equation_COG_1}\\ 
\end{equation}
where $N$ is the total number of detected PGs during an acquisition run; $N_{Det}$ is the number of pixels composing the detector; $\textbf{r}_{i}$   represents the coordinate vector of the $i^{\mathrm{th}}$ pixel;
and  $n_{i}$ is the number of PGs detected by the  $i^{\mathrm{th}}$ pixel.\\
Since TIARA pixels cover the target in 3D, by definition, \textit{y$_{COG}$} and \textit{z$_{COG}$} are correlated to the beam position in the transverse plane, whereas \textit{x$_{COG}$} provides a parameter associated to the average beam penetration in the target. 
As the number of PGs detected by a pixel essentially depends on the solid angle subtended by the pixel itself and the beam, any displacement of the beam in the lateral plane or along the distal direction  produces, in principle, a measurable deviation of the corresponding(s) COG value(s).\\
It should be noted that, while the sensitivity of the PG vertex reconstruction method described in section~\ref{AR_section} directly depends on the detection system time resolution, in the COG  method, the PG TOF is not used for signal reconstruction. It will be shown in the next sections that a very limited time resolution is only needed for background rejection. 
\subsubsection{COG sensitivity.}
The sensitivity of the COG-based reconstruction algorithm was evaluated through MC simulations for both a lateral beam displacement and  proton range shift along the beam direction. 
A comparison between treatment and treatment planning was performed for a \textit{monitoring statistics} of $10^7$ or $10^8$ incident protons, supposedly corresponding to a real-time monitoring procedure performed within the first irradiation spot(s). \\\\
\textit{Simulations of range deviations}\\
The COG method capability to detect a distal proton range shift and a lateral beam displacement were studied separately.
In the first case, data from the MC simulations including an air cavity of variable thickness (see section \ref{cavity}) were reconstructed using equation~\ref{equation_COG_1} to calculate \textit{x$_{COG}$}.
In the second case, the uniform head geometry implemented for the validation of the analytical reconstruction algorithm (figure~\ref{tiara_simulation},~left) was considered. For this simulation, the incident beam was displaced in the $y$-axis direction in progressive steps of 1 mm (from 0 to 5 mm), and the corresponding \textit{y$_{COG}$} was calculated. Because of the cylindrical symmetry, a displacement along the $y$-axis has the same impact as a displacement along the $z$-axis. \\\\
\textit{COG sensitivity determination}\\
In principle, the COG algorithm sensitivity may be estimated through the simulation of repeated experiments as it was done for the analytical reconstruction, but this strategy would be excessively time-consuming in this case, as statistically independent subsets are needed.
In the previous analysis, the event-by-event time convolution guaranteed the statistical independence of the 10$^7$/10$^8$ protons subsets built from the 1.5 $\times$ 10$^9$ simulated events, since the particle TOF was the physical quantity used for the reconstruction.  Conversely, the COG calculation  mainly depends on the number of PGs detected in each pixel, which only slightly varies after the time- and energy-convolutions are performed  and the \textit{Tcut} and \textit{Ecut} are applied. Therefore, a different approach was used to generate the simulated experiments.
For each displacement simulated on the \textit{x}- and \textit{y}-axis, the TPS statistics of  $1.5 \times 10^{9}$ incident protons was initially considered. The TOF and energy distributions obtained after convolution with realistic time- and energy-resolutions, were used to calculate the average number of hits in each detector ($n_i$).  Given the high proton statistics used, it could be safely assumed that the ratio $n_i$/$N$ converges, and thus $n_i$  could be interpreted as the mean ($\lambda_{i,TPS}$) of a Poisson distribution describing the hits in the  $i^{\mathrm{th}}$ pixel at \textit{TPS statistics}. \\
In order to build 5$\times$10$^{4}$ simulated experiments at \textit{monitoring statistics} (10$^7$ or 10$^8$ incident protons), the $\lambda_{i,TPS}$ associated to each detector pixel was respectively divided by a factor 150 and 15 (e.g. the latter corresponds to $1.5 \times 10^{9} /10^{8} $). Then, the $\lambda_i$ values obtained were used to build two different collections of normalised pdfs using a Poisson-distribution-based MC approach: the first corresponding to the six air cavity thicknesses (distal proton range shift); the second to the  six lateral beam displacements. From these pdf collections, the type-II errors associated to different distal and lateral beam variations were calculated as described in section \ref{cavity}.
\section{Results}
\subsection{PG vertex reconstruction}
\subsubsection{Vertex reconstruction validation.}
\begin{figure}[!h]
    \centering
    \includegraphics[width =0.45\textwidth]{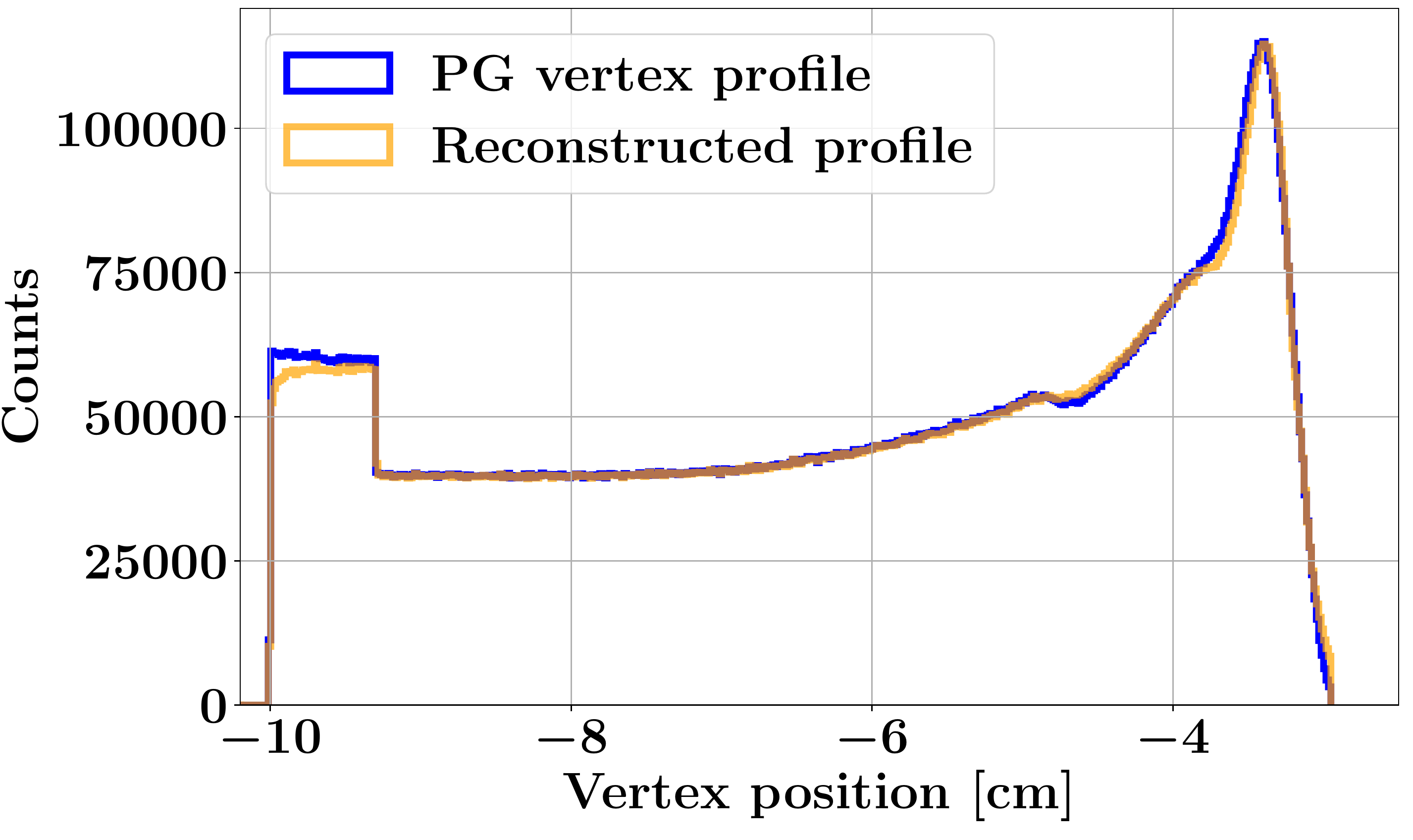} 
       \includegraphics[width =0.45\textwidth]{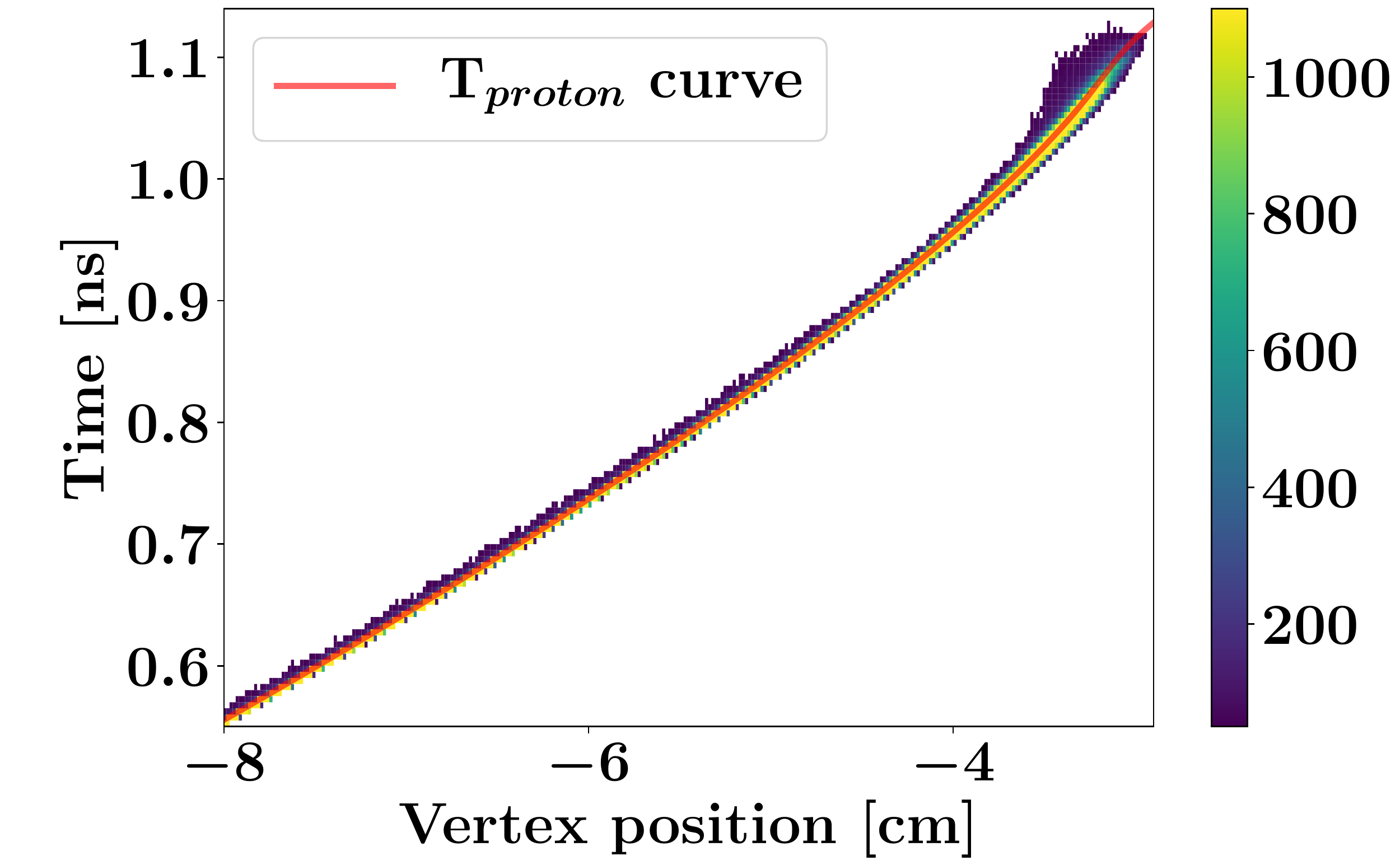} 
\caption{Left: comparison
    between the reconstructed PG-vertex  profile  and the distribution of  PG vertex coordinates directly issued from the Geant4 simulation using a 100 MeV point-like proton beam. A data selection above 2 MeV has been applied in the reconstruction to remove most of the scattered PGs.  Right:  emission time of the PG recorded in the phase space, as a function of the proton beam penetration. The red line is the simulated $T_{proton}$ curve corresponding to the average time at which the impinging protons reach a given depth in the target. 
    }
\label{RC_PG}
\end{figure}

Figure~\ref{RC_PG},~left shows the comparison of the reconstructed and simulated PG vertex profiles recorded in the phase space (i.e at perfect time or energy resolutions). The first profile only includes  the vertices of the unscattered primary PGs and is obtained using the $T_{proton}$ curve shown (in red) in figure~\ref{RC_PG},~right to resolve equation~\ref{eqfinale};
the second one is built using the PG vertices directly retrieved from simulated data and corresponding to the 2D histogram in figure~\ref{RC_PG},~right.
The two profiles of figure~\ref{RC_PG},~left are quasi-identical over almost the whole proton range, thus validating  the proposed reconstruction algorithm in one dimension.\\
However, two main discrepancies are observable: the first in the  (-10; -9.3) cm region, corresponding to the skull; the second one going from the PG profile peak to the fall-off. \\
\noindent In the skull, the lack of reconstructed PG is related to the $^{40}$Ca(p,p')$^{40}$Ca$^*$  reaction, generating PGs of 3.736 MeV with a mean lifetime of 47 ps (for a 100 MeV proton this corresponds to a mean path of $\sim$6~mm). For those emissions, the hypothesis of $T_{decay}=0$, which is the cornerstone of time-based PG monitoring approaches, no longer stands and the corresponding reconstructed events are spread out at larger penetration depths.\\
In the peak region, a shift of approximately 1 mm between the two profiles  is clearly identifiable. This effect is mainly associated to the  $^{16}$O(p,p')$^{16}$O* reaction emitting 6.13 MeV gamma rays with a mean lifetime of 27 ps, and whose cross section is peaked at the end of the proton range. According to their actual emission position and time, the vertex of these PG may be reconstructed at larger depths or even outside the profile.
Nonetheless, although the agreement between the actual and the reconstructed profiles is not perfect, it should be noticed that this level of inaccuracy is much smaller than the time resolution achievable by a gamma-ray detector. \\
Some comments should also be made on the non-physical discontinuity observable at $\approx$-5~cm in both the reconstructed and  simulated profiles. This anomaly is caused by an abrupt change in the Geant4 physical simulation processes (i.e Fermi break-up activation) triggered by the decrease of the proton energies below 45 MeV, and leading to a decrease of nuclear fragment excitation energy (mainly for carbon and oxygen). The overall effect is a decrease in the PG production rate of these nuclei (Verburg \etal 2012).
\subsubsection{Reconstructed PG profiles for the reference irradiation.}
 \begin{figure}
    \centering
    \includegraphics[width = 0.6\textwidth]{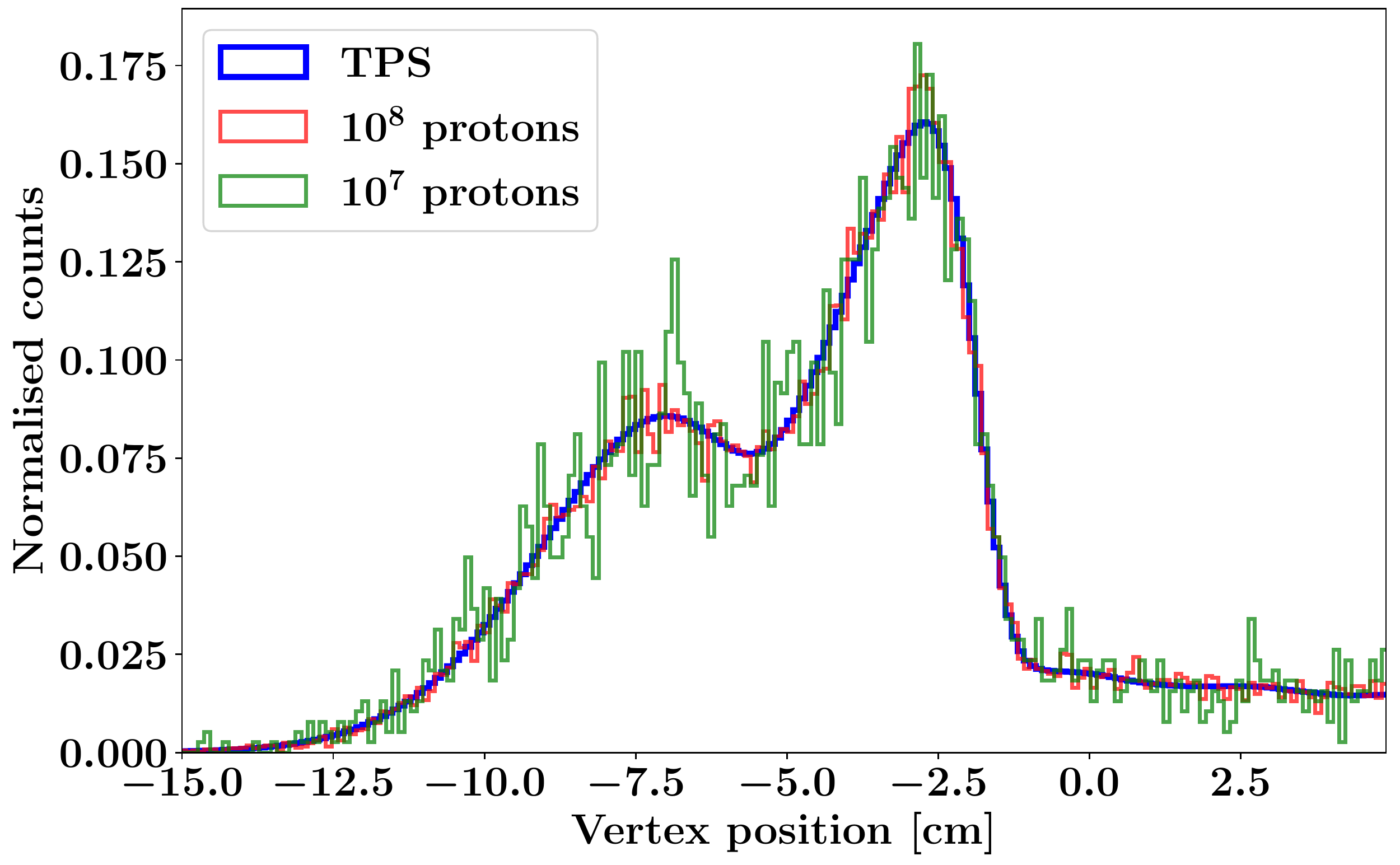}
    \caption{Reconstructed PG profiles for the 1 cm thick air cavity geometry obtained with different \textit{monitoring statistics} and at \textit{TPS statistic} (1.5$\times$10$^9$ incident protons). The \textit{Ecut} was applied to all datasets.  }
    \label{rc_stats}
\end{figure}
\begin{figure}
    \centering
    \includegraphics[width = 0.7\textwidth]{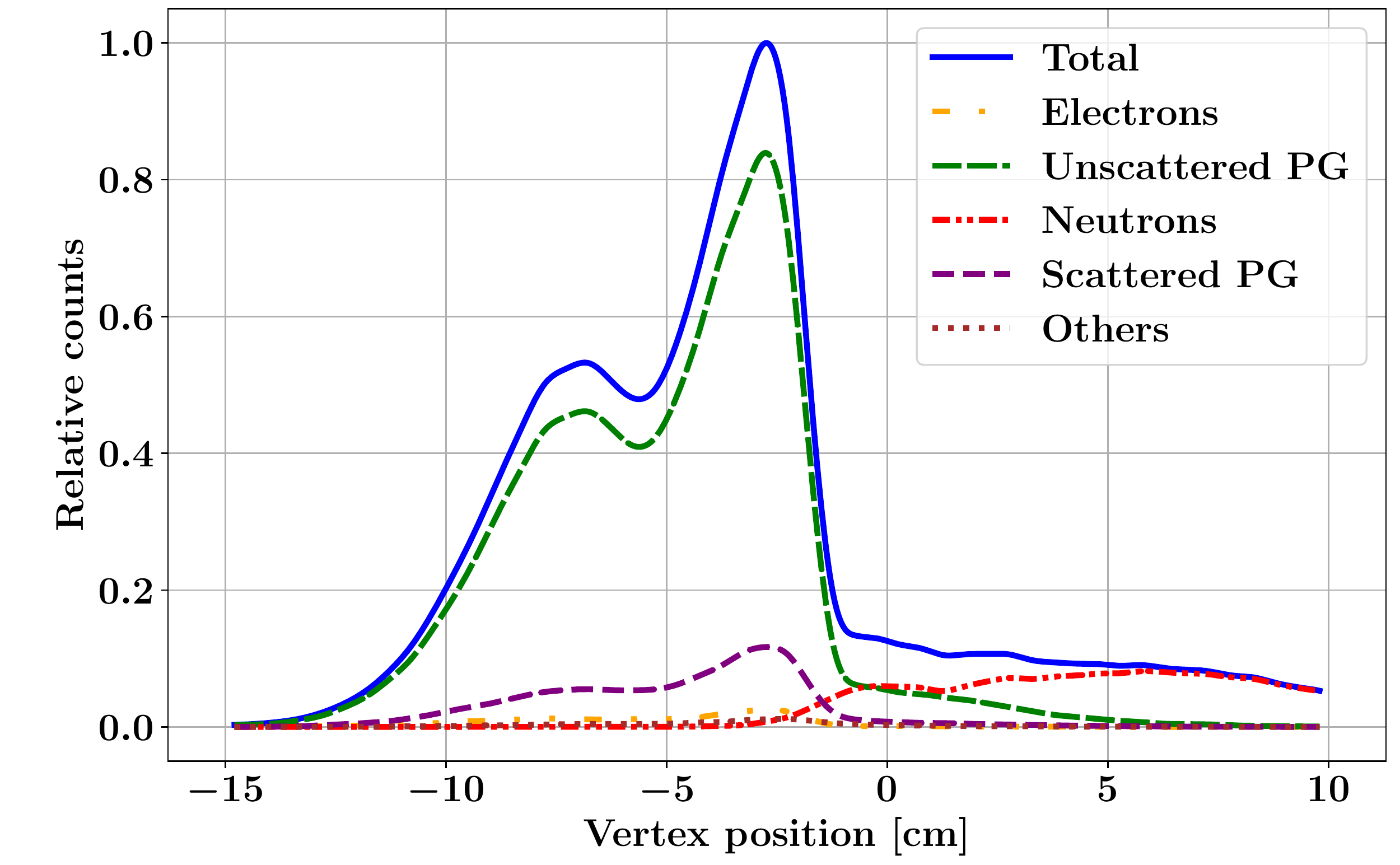}
    \caption{The reconstructed PG vertex profile obtained at \textit{TPS statistics} and shown in figure~\ref{rc_stats}  after the \textit{Ecut} was applied, is decomposed to show
 the separate contributions of the different secondary particle types detected in the phase space.  Slower particles as neutrons (red curve) are naturally time-discriminated from the other particles. }
    \label{rc_particle_types}
\end{figure}
\begin{figure}
    \centering
    \includegraphics[width = \textwidth]{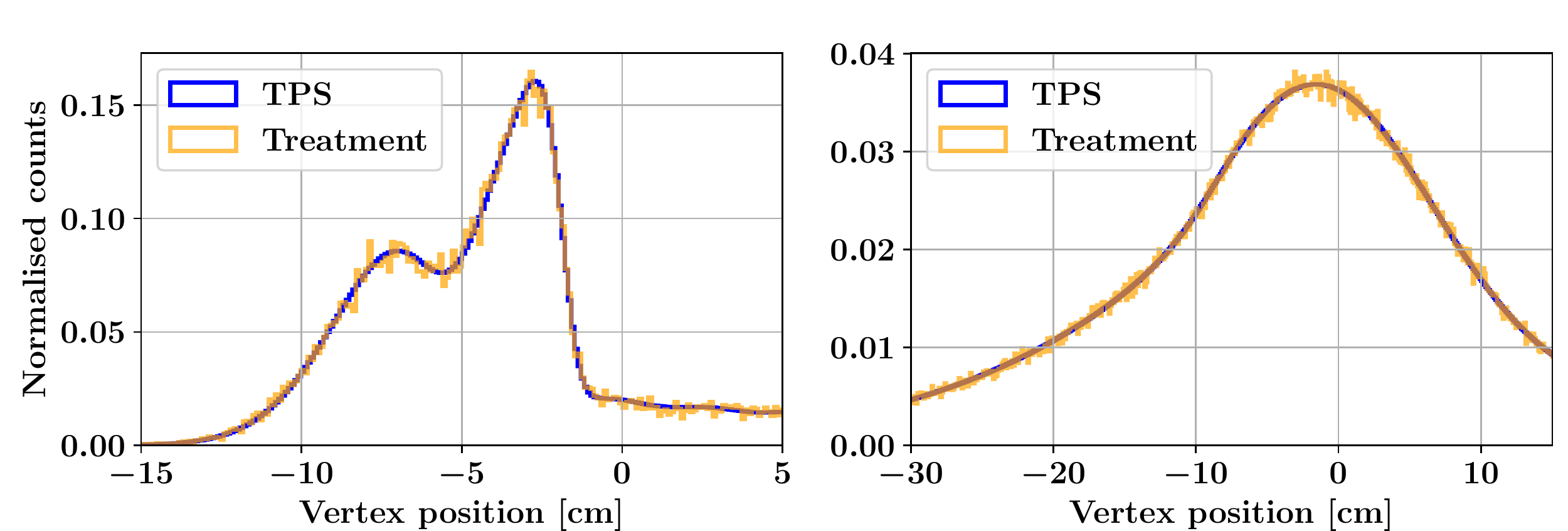}
    \caption{Comparison of PG vertex profiles reconstructed in TPS (blue curve) and treatment (orange curve) conditions in the case of a 1 cm thick air cavity. In both graphs, a high statistics of 1.5 $\times$ 10$^{9}$  incident protons was employed for the TPS profile. On the left, the treatment profile is reconstructed assuming a 100 ps (rms) CTR and a statistics of 10$^8$ incident protons, corresponding to $\approx$ 3$\times$10$^4$ PG detected, whereas on the right a CTR of 1 ns (rms) was considered, with a statistics of 10$^9$ incident protons, corresponding to $\approx$ 3$\times$10$^5$ PG detected.
   The \textit{Ecut} (E > 2 MeV) was applied to data in both  the left and right graphs.}
    \label{comparison_HS_BS}
\end{figure}

The PG vertex profiles obtained with the geometry including a 1 cm air cavity  described in section \ref{cavity} are presented in figure~\ref{rc_stats} for the different \textit{monitoring statistics} considered (10$^7$ and 10$^8$ incident protons) and the \textit{TPS statistics} corresponding to 1.5$\times$10$^9$ incident protons. The profiles are obtained after convolution of the simulated data with a CTR of 100 ps (rms) and an energy resolution of 1 MeV (rms). Before normalisation to unit area, and after the \textit{Ecut} is applied, approximately 3700 and 36600 PG events are included in the two histograms at \textit{monitoring statistics}, and about 548000 events in the histogram at \textit{TPS statistics}. From a qualitative point of view, all profiles, including those at lower statistics, clearly show the presence of the air cavity, and a quite sharp fall-off after the maximum. \\
All  distributions also present a distinct tail after the profile  fall-off. In order to better understand the origin of this tail, figure~\ref{rc_particle_types} shows the separate  contribution of the different secondary particles detected by TIARA in the case of the PG profile obtained at \textit{TPS statistics} and already presented in figure~\ref{rc_stats}.
  This graph confirms that the \textit{Tcut} is not necessary to reject neutrons correlated to the beam; this may not be true for experimental data including neutrons from ambient noise. 
 Nevertheless, it should be noticed that the neutron contribution is overestimated in these simulations, since the same detection efficiency was considered for all particles. A pure Cherenkov radiator will most probably detect fewer neutrons (both the time-correlated and the ambient neutrons), therefore decreasing the amplitude of the associated tail.  \\
Finally, in figure~\ref{comparison_HS_BS}, only the 1 cm cavity profiles corresponding to the \textit{TPS statistics} and a \textit{monitoring statistics} of 10$^8$ incident protons have been selected to present the impact of system CTR  on the PG profiles shape: in the left graph, simulated data are convolved with a time resolution of 100 ps (rms), whereas a time resolution of 1 ns (rms) was assumed for data shown on the right. In both cases, an energy resolution of 1 MeV (rms) was considered, and the \textit{Ecut} was applied. Again, the presence of the  1 cm air cavity is clearly identifiable on the reconstructed profiles at 100 ps resolution (left) for both the high and the low statistics considered, whereas the cavity is not detectable on the 1 ns resolution data.\\
\subsubsection{PG vertex reconstruction sensitivity.}
\begin{figure}[!h]
    \centering
    \includegraphics[width = \textwidth]{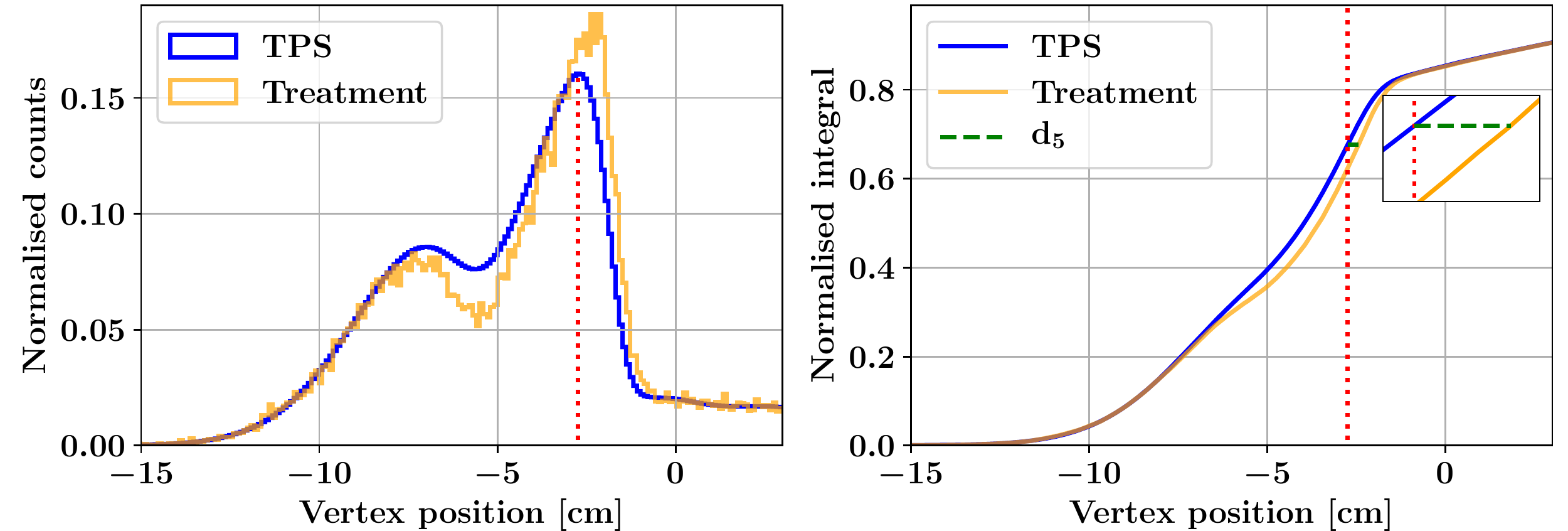} 
    \caption{Comparison of PG vertex profiles (left) and integrated PG vertex profile (right)
    obtained for the reference simulation in  TPS conditions (1 cm air cavity) (blue)  and for a  treatment  geometry (orange) presenting a 1.5 cm thick  air cavity. In both plots, the vertical dotted line (in red) locates the TPS PG profile peak at $x_{TPS}$. In the right plot, the horizontal dashed line (in green) in the inset represents the calculated profile distance $d_{5}$.}
    \label{comparaison_profil_integre}
\end{figure}
\begin{figure}[!h] 
\centering 
\includegraphics[width = \textwidth]{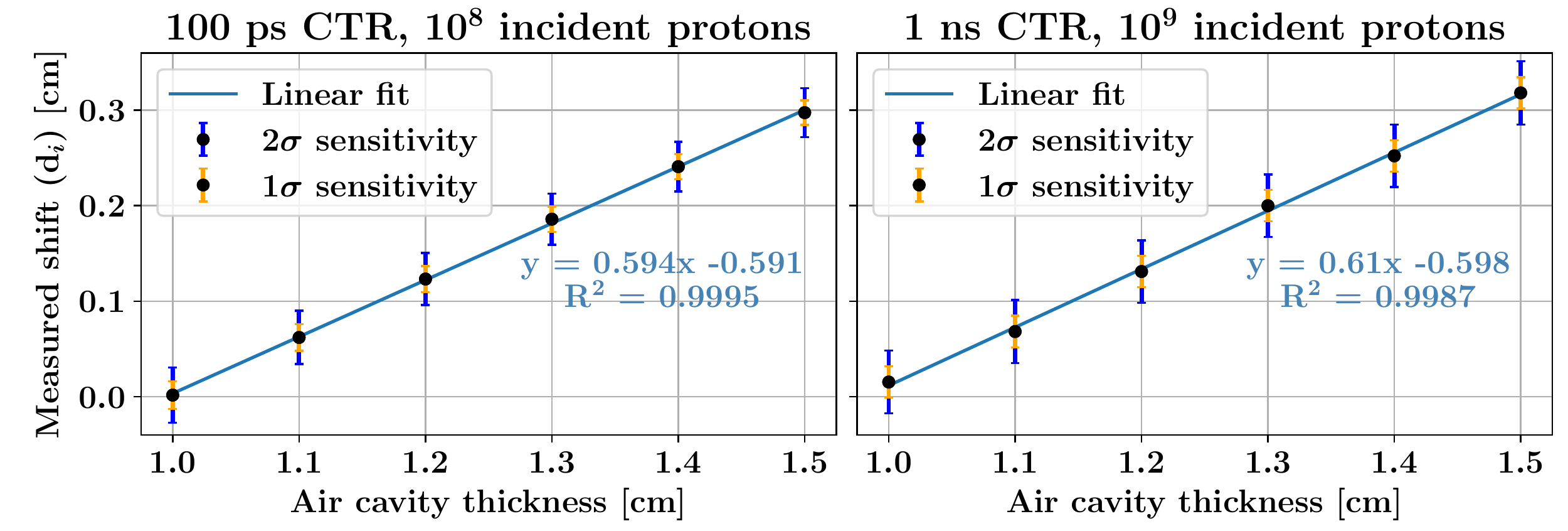}
\caption{Correlation between the actual air cavity thickness and the corresponding measured profile distance $d_{i}$ in the case of a 100 ps (rms) CTR and 10$^8$ incident protons (left), and in the case of a 1 ns (rms) CTR and a statistics of 10$^9$ incident protons (right). The 
1$\sigma$ (orange) and 2$\sigma$ (blue) errors are displayed for each data point. A linear fit was carried out for each dataset.} \label{shift_measurement_correlation}
\end{figure}

A comparison between the reference PG profile (1 cm thick air
cavity at 1.5$\times$10$^9$ incident protons) and the treatment PG profiles (air cavity thickness ranging from 1.0 to 1.5 cm) was carried out for both a CTR of 100 ps (10$^8$ incident protons) and 1 ns (10$^9$ incident protons), with the aim of estimating the sensitivity of the analytical reconstruction.
Before normalisation to unit integral, each PG profile roughly includes 3$\times$10$^{-4}$  PGs per incident proton. 
In figure~\ref{comparaison_profil_integre},~left, the reconstructed reference (TPS) profile and the treatment profile corresponding to the 1.5 cm air cavity thickness
are presented in the case of a 100 ps time resolution. As expected, the two curves display no difference at beam entrance and before the air cavity, whereas the larger discrepancy is found in the PG profile peak region. It may be noted that, the fall-off of the treatment profile is sharper than one would  expect. This effect depends on the use of a $T_{proton}$ function determined on the TPS geometry (1 cm thick cavity) to reconstruct  treatment data  (1.5 cm thick cavity in this case). It should therefore be stressed that, with this approach, the treatment's reconstructed profile does not correspond  to the real PG vertex distribution in presence of an anatomical variation. However,  it will be shown in the next sections that it contains enough information to allow the detection of a discrepancy between treatment and treatment plan.\\
Figure~\ref{comparaison_profil_integre},~right, shows the integrated PG profiles used to calculate the profile distance $d_5$ according to the method described in section \ref{cavity}: here, $d_5$ corresponds to the green horizontal dashed line shown in the inset. The same analysis was performed for all cavity thicknesses  and the two time resolutions considered. Data are summarised in figure~\ref{shift_measurement_correlation} for the 100 ps (left) and 1 ns (right) time resolutions. The graphs display the excellent correlation between the average profile distance $d_i$ measured from the pdfs of the experiments and the actual range shift induced by the variable cavity thickness. Error bars corresponds to the 1$\sigma$ (orange) and 2$\sigma$ (blue) statistical uncertainties obtained from the normalised pdfs of $d_i$ which are built from the simulated experiments. The strong linearity of these data and the limited statistical errors
suggest that a millimetric precision  on the
 distal proton range shift could be achievable in both scenarios considered. 
More quantitatively, the analysis of type-II errors obtained from the pdfs for $d_i$ confirms that a
1 mm distal proton range shift would be detectable at 2$\sigma$ with a 100 ps time resolution within one intense irradiation spot of 
10$^{8}$ protons. At the same time, a 2 mm range shift seems detectable at 2$\sigma$ in the 1 ns CTR scenario, when employing 10$^9$ incident protons for the monitoring procedure. 
The 1$\sigma$ and 2$\sigma$ sensitivities obtained for the different combinations of \textit{monitoring statistics} and CTRs considered in this study are summarised in table \ref{sens_summary}. 
\subsection{Centre of gravity}
\subsubsection{COG sensitivity to lateral beam displacements.} 
\begin{figure}[!h] 
\centering 
\includegraphics[width = \textwidth]{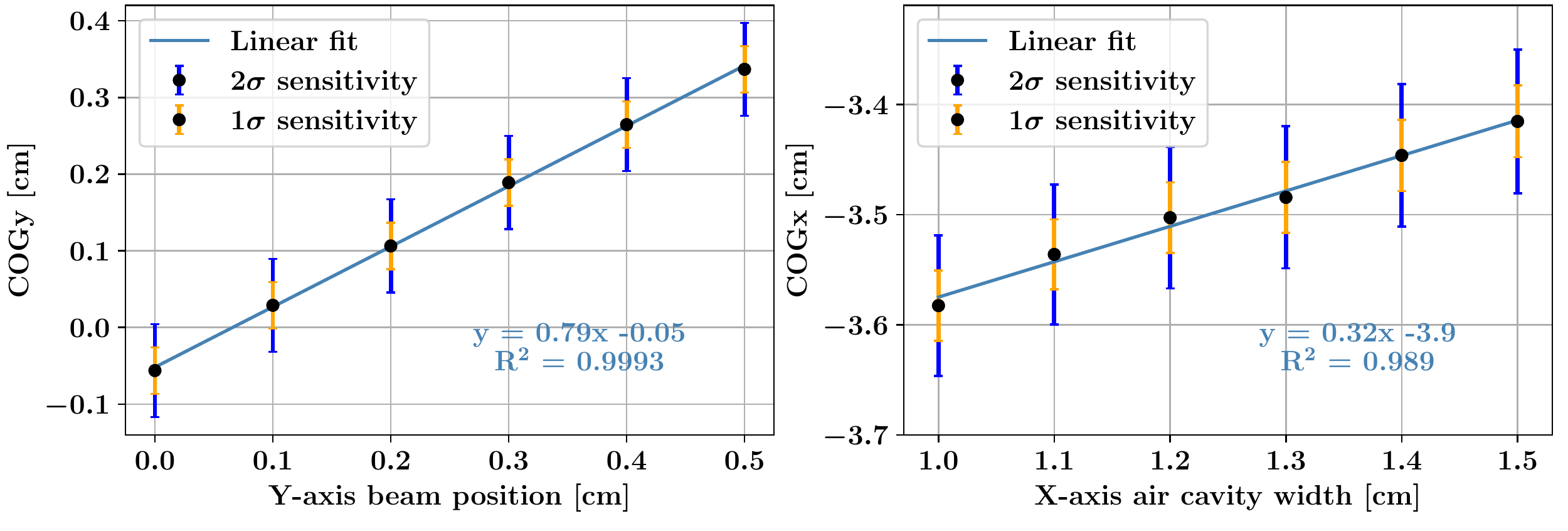}
\caption{Left: correlation between the detectors \textit{y$_{COG}$} calculated under treatment conditions and the actual beam displacement imposed in the MC simulation. Right: correlation between the calculated \textit{x$_{COG}$} and the actual cavity thickness considered in the MC simulation. Both datasets correspond to a \textit{monitoring statistics} of 10$^{8}$ incident protons. The 1$\sigma$ (orange) and 2$\sigma$  (blue) errors are displayed in both graphs.  A linear fit was carried out for each dataset.
} 
\label{COG_correlation} 
\end{figure}
Figure~\ref{COG_correlation},~left shows the correlation between the calculated \textit{y$_{COG}$}  and the lateral beam displacement imposed in the MC simulation. Error bars correspond to the 1$\sigma$ (orange) and the 2$\sigma$ (blue) statistical errors computed from the simulated experiment pdfs. Both the \textit{Ecut} and the \textit{Tcut} were applied to these data.
From this plot, it can  be observed that the linear correlation functions do not cross the  origin of the coordinate system. 
This effect comes from the fact that TIARA pixels locations are uniform on the phase space spherical surface, but they are not symmetric with respect to the beam axis.
Nevertheless, since the correlation between the calculated COG and the actual beam displacement is linear, a calibration procedure could be envisaged before treatment.\\
From the type-II error analysis, a 2 mm sensitivity was found at 2$\sigma$ for 10$^{8}$ incident protons, while the sensitivity obtained at 1$\sigma$ is 1 mm. 
\subsubsection{COG sensitivity to distal proton range shifts.}
While the COG method seems very promising for the detection of  lateral proton beam displacement, its application to the detection of distal shifts seems less relevant with TIARA design. Figure~\ref{COG_correlation},~right, shows that the correlation between the calculated \textit{x$_{COG}$} and the actual air cavity thickness can still be fitted with a linear function, but its slope is much smaller than the one found for the \textit{y$_{COG}$}. Since the error bars are of the same order of magnitude as those found for \textit{y$_{COG}$}, the limited slope directly translates into a loss of sensitivity. 
From data presented in figure~\ref{COG_correlation},~right, a 4 mm proton range shift seems detectable  at 2$\sigma$ with 10$^{8}$ incident protons, which is 3-4 times larger than the shift detectable using the analytical reconstruction. At the same time, a 2 mm sensitivity was found at 1$\sigma$.\\
The 1$\sigma$ and 2$\sigma$ sensitivities obtained for the different \textit{monitoring statistics} considered in the COG approach, are summarised in table \ref{sens_summary}. 
\section{Discussion}
With the goal of developing a multi-channel detector (TIARA) for proton range monitoring, we propose a dedicated reconstruction algorithm that has the potential to provide 3D information on the PG vertex distribution. The algorithm is non-recursive and may therefore be implemented to obtain a real-time response during the monitoring procedure.
The originality of this approach consists in  unfolding the PG TOF from the overall (proton plus PG) time measurement, which is the key for combining the response of multiple PGT detectors  within a single acquisition and thus improving the detector sensitivity to a possible beam displacement.  In this way, a full 3D coverage can be achieved, that allows detecting an irradiation deviation in any direction. In this sense, TIARA can also be interpreted as a multi-channel ($\sim$30) position-sensitive detector, with a limited angular coverage and a spatial resolution given by the pixel extension. \\
The proposed reconstruction formula (see equation~\ref{eq_TIARA_tmp_2}) describes the PG vertex distribution in the three dimensions, but its solution is non-trivial as it requires the resolution of an inverse problem (e.g. by machine learning, Bayesian approaches, etc.). 
In the present work, equation~\ref{eq_TIARA_tmp_2} was only resolved in one dimension, to obtain the longitudinal PG vertex distribution. This approximation corresponds to having an  \textit{a-priori} knowledge of the proton trajectory, as it would be possible, for example, using a position-sensitive beam monitor (Gallin-Martel \etal 2018). \\
Besides, an additional reconstruction method based on the COG calculation was also implemented, that  adapts the PGPI technique (Krimmer \etal 2017) exploiting the 
knowledge of the PG hit coordinates, an information available when using small pixels to detect gamma rays (i.e position-sensitive detectors) as in TIARA. This approach is complementary to the vertex reconstruction algorithm as it weakly depends on the system time resolution and can be applied without the need for reducing the beam intensity. In this work, it also allowed us to go beyond the 1D approximation and preliminarily explore the sensitivity that could be achieved by TIARA in the transverse plane. Most probably, the COG approach would became redundant once a 3D  analytical reconstruction will be implemented, as the physical variables included in the COG reconstruction are a subset of those necessary for the analytical reconstruction. In any case, 
being able to observe a possible lateral beam shift only using data from the gamma detector would be especially convenient, as it would eliminate the necessity of measuring the proton incident position, thus reducing the beam monitor electronic channels to a single one (dedicated to the measurement of time).\\
\begin{table}[t]
\caption{\label{sens_summary} Summary of the 1$\sigma$ and 2$\sigma$ sensitivities obtained with the different reconstruction methods and the two CTRs considered. VR stands for Vertex Reconstruction.}
\begin{indented}
\item[]\begin{tabular}{@{\extracolsep{4pt}}lccccccc@{}}
\br
        &   \multicolumn{5}{c}{\textbf{Distal proton range shift}}& \multicolumn{2}{c}{\textbf{Lateral beam shift}}\\
         Method & \multicolumn{2}{c}{100 ps VR}  & 1 ns VR & \multicolumn{2}{c}{\textit{x$_{COG}$}} & \centre{2}{\textit{y$_{COG}$}}\\
          \cline{2-3} \cline{4-4} \cline{5-6} \cline{7-8}
    Nb of protons& $10^{7}$ & $10^{8}$ & $10^{9}$ & $10^{7}$ & $10^8$  &  \hspace{0.6 cm}$10^{7}$ & $10^{8}$\\
\mr     
1$\sigma$ sensitivity (mm) & 2 & 1 &    1  &  >5 & 2  &  \hspace{0.6 cm} >5 & 1\\
2$\sigma$ sensitivity (mm) & 3 & 1 &    2  &  >5 &    4  &  \hspace{0.6 cm} >5 & 2\\
\br
\end{tabular}
\end{indented}
\end{table}
Although the reconstruction strategies developed so far are still sub-optimal, a MC-based sensitivity study has allowed demonstrating that a detector with the TIARA characteristics would be able to detect a difference between treatment and treatment plan with a  precision of a few mm at pencil beam scale (10$^7$ -- 10$^8$ protons). Different scenarios were investigated in this work and the results are summarised in Table \ref{sens_summary}. The sensitivity of the PG vertex reconstruction approach strongly depends on the detection system CTR. A 100 ps (rms) CTR can be realistically obtained employing fast gamma ray and beam-monitor detectors in single proton regime  (Marcatili \etal 2020). Under this hypothesis, and for the simplified geometry considered, we have proven that a distal proton range shift  of $\sim$ 1 mm would be detectable at 2$\sigma$ with  10$^{8}$ incident protons, roughly corresponding to a single intense irradiation spot (Smeets \etal 2012). A slightly worst sensitivity of 3 mm (at 2$\sigma$) has been found for  10$^{7}$ incident protons.
The rest of the treatment could be carried out at the nominal beam intensity, and TIARA data acquisition could continue with degraded timing performances. In this case, the time resolution would be basically dominated by the proton bunch time-width ($\approx$ 1 ns rms for a C230 IBA cyclotron)  that, in turn, depends on the beamline settings and the requested proton energy. Here, a worst case scenario of a 1 ns (rms) CTR was considered to show that the limited CTR can be compensated by the acquisition of a higher statistics: a 2 mm range shift can still be detected at 2$\sigma$ under these conditions, when considering  10$^{9}$ incident protons. In this case it could be possible to combine  a group of neighboring spots (Xie \etal 2017). \\
On the other hand, the COG approach only weakly depends on the system time resolution (for background rejection), and could be applied during the whole treatment at clinical beam currents (without intensity reduction). This method was proven advantageous for the detection of lateral beam displacements (a 2 mm shift is detectable with  10$^{8}$ incident protons), but its use for the detection of range shifts along the beam direction seems less promising with  TIARA geometrical configuration. \\
It is important to clarify that the results listed in Table \ref{sens_summary} are indicative and should be interpreted as orders of magnitude, since the simulated sensitivity may depend on a number of parameters 
whose impact was not evaluated in the current MC study. These include the finite extension of the proton beam, its energy dispersion, the 
presence and localisation of multiple heterogeneities in the target and the PG yield considered in the MC simulation.
 This last parameter is known to  vary among different physics lists and also between different versions of the same physics list (Arce \etal 2020, Pinto \etal 2016), with values that can be relatively distant from those experimentally available. In this paper we have considered the QGSP\_BIC\_HP\_EMY physics list, which is recommended by Geant4 developers for hadrontherapy simulations, and the version 10.4.patch02 of the Geant4 toolkit, obtaining a quite high PG yield of ~0.09 PGs per incident 100 MeV proton that positively biased the method sensitivity. For example, we have obtained a PG yield of 0.09 and 0.07 when using Geant4 version 10.04 and 10.06 (the newest one) respectively, but we have selected version 10.4.patch02 because it offered a more physically realistic  energy distribution of PG rays.\\
In addition, all MC simulations described in this paper were implemented using a detection phase space: the detector response was reproduced \textit{a posteriori}, using the approach described in section \ref{det_resp}.  This procedure allows to easily associate a number of detected PGs per incident protons for a specific detector configuration, making it possible to investigate TIARA sensitivity. Nevertheless, its main limitation is to impose the same detection efficiency to all particle types. On one hand, neutron detection probability is expected to be much lower than  26.6\% in a pure Cherenkov radiator. Thus, the simulated data include an excess of detected neutron,  going in the direction of  an  underestimation of the system Signal-to-Noise Ratio (SNR). On the other hand, the PG detection efficiency is not expected to be constant in the PG energy range. In particular, in a real experiment, the detection of the low-energy scattered PG component would be more favorable: the direct consequence is an overestimation of the SNR at low energies in the simulated data. 
It is clear therefore, that the actual potential of TIARA
should ultimately be established through experiments.\\
One element of originality of the proposed PG detector consists in the prospective use of small-size Cherenkov radiators for gamma detection, with the aim of pushing  the limits of the system time resolution. Indeed, the Cherenkov process is inherently much faster than the scintillation one, and the high stopping power of many radiators, such as PbF$_{2}$, grants the use of crystals with an high aspect ratio (Gundacker \etal 2014), favouring the time resolution without sacrificing the detection efficiency.  The drawback is the very poor energy resolution of the PG detector because the limited pixel size does not allow to fully absorb the energetic PGs.
For this reason, a very poor, arbitrary energy resolution of 1 MeV (rms) was considered in this paper. 
Since the unscattered PG energy spectrum (cf. figure~\ref{selection_criteria}, green curve) is reduced to an poorly defined bump after the 1 MeV energy convolution, 
we believe this value reasonably corresponds to a worst case scenario.
The excellent proton range shift sensitivity obtained despite this assumption is possible because, for TIARA, the energy information only plays  a role in background rejection and not in  signal reconstruction. 
Clearly, simulated data do not include scattered PGs from the environment outside the patient; however, in a very preliminary experiment (Marcatili \etal 2019), we have shown that a clean PGT spectrum could be acquired with a non-optimised gamma detector composed of a 5$\times$5$\times$20 mm$^{3}$ PbF$_{2}$ crystal, read out by a 1$\times$1 mm$^{2}$ RGB-SiPM from Hamamatsu and using a slow preamplifier. With an estimated CTR of $\approx$300 ps (rms), it was possible to measure the penetration of a 68 MeV proton beam in a block of PMMA, placing the detector  at  few centimeters distance from the target.
Despite the low SiPM detection area and the non-optimised optical coupling, up to 8-9 photoelectrons per incident PG could be detected; an energy threshold between 3 and 4 photoelectrons was applied.\\
We have recently started a research project to carry out TIARA R\&D at the pixel level, and to find the optimum crystal and photodetector candidates. Within the same project, a fully 3D PG vertex reconstruction algorithm will be also developed.\\
Finally, the potential performances of the proposed approach were only investigated in the framework of protontherapy. 
Equation~\ref{eq_TIARA_tmp_2} remains valid in the case of heavier ions, but the impact of secondary protons, whose vertices lay outside  the beam-path, remains to be investigated in both the 1D approximation as well as in the fully-3D solution. It can be anticipated, however, that  a beam current reduction would not be necessary for $^{12}$C treatments,  for which 2-5 ions/bunch are typically delivered in 20-50 ns bunches (Dauvergne \etal 2020) at European synchrotrons; as a consequence, TIARA could be operated in high CTR mode for as many  irradiation spots as needed.
At the same time, better system CTRs are achievable in carbontherapy because the  higher Linear Energy Transfer (LET) of $^{12}$C results in larger deposited energies in the beam monitor, leading to better timing performances.
\section{Conclusion}
This work introduces a novel detection system for online range monitoring in protontherapy based on PG Time Imaging (PGTI), through the development of a dedicated reconstruction algorithm. 
The proposed detector has the potential of providing 3D information on proton range with millimetric precision, with a simple (few independent electronic channels) and cheap (large-size, expensive, scintillating crystals are not required) detector design. It can be either employed to monitor the treatment in real-time (within the first or first few irradiation spots), if a reduction of the beam current is feasible, or within one treatment session if a degraded CTR is accepted to conform to the nominal beam current. A research project has recently started with the aim of developing a fully-3D reconstruction algorithm and of defining the TIARA pixel design.
\section{Acknowledgements}
This work was partially supported by the ANR project ANR-15-IDEX-02 and by INSERM Cancer.
This work was performed within the framework of the LABEX PRIMES (ANR-11-LABX-0063) of Universit\'{e} de Lyon.
\section*{References}
\begin{harvard}
\item{} Agostinelli S et al. 2003 GEANT4 simulation toolkit, \textit{Nuclear Inst. and Methods in Physics Research A} \textbf{506.3} 250-303. \href{https://doi.org/10.1016/S0168-9002(03)01368-8}{doi:10.1016/S0168-9002(03)01368-8}
\item{} Arce P et al. 2020 Report on G4-Med a Geant4 benchmarking system for medical physics applications developed by the Geant4 Medical Simulation Benchmarking Group, \textit{Med. Phys.} Accepted Author Manuscript. 
\href{ https://doi.org/10.1002/mp.14226}{doi:10.1002/mp.14226} %
\item{} Bisogni MG et al. 2016 INSIDE in-beam positron emission tomography system for particle range monitoring in hadrontherapy, \textit{J. Med. Imaging} \textbf{4(1)} 011005.
\href{https://doi.org/10.1117/1.JMI.4.1.011005}{doi:10.1117/1.JMI.4.1.011005}
\item{} Dauvergne D et al. 2020 On the role of single particle irradiation and fast timing for efficient online-control in particle therapy, \textit{Front. Phys.} \textbf{8} 434. 
\href{https://doi.org/10.3389/fphy.2020.567215}{doi:10.3389/fphy.2020.567215}
\item{} Dosanjh M et al. 2018 ENLIGHT: European network for Light ion hadron therapy, \textit{ Radiother. Oncol.} \textbf{128.1} 76-82. 
\href{https://doi.org/10.1016/j.radonc.2018.03.014}{doi:10.1016/j.radonc.2018.03.014}
\item Draeger E et al. 2018 3D prompt gamma imaging for proton beam range verification, \textit{Phys. Med. Biol.} \textbf{63.3} 035019.
\href{https://doi.org/10.1088/1361-6560/aaa203}{doi:10.1088/1361-6560/aaa203}
\item{} Enghardt W et al. 2004 Dose quantification from in-beam positron emission tomography, \textit{Radiother. Oncol.} \textbf{73} S96–8.
\href{https://doi.org/10.1016/S0167-8140(04)80024-0}{doi:10.1016/S0167-8140(04)80024-0}
\item{} Ferrero V et al. 2018 Online proton therapy monitoring: clinical test of a Silicon-photodetector-based in-beam PET, \textit{Sci. Rep.} \textbf{8.1} 4100.
\href{https://doi.org/10.1038/s41598-018-22325-6}{doi:10.1038/s41598-018-22325-6}
\item{} Gallin-Martel ML et al. 2018 A large area diamond-based beam tagging hodoscope for ion therapy monitoring, \textit{EPJ Web Conf.} \textbf{170} 09005. 
\href{https://doi.org/10.1051/epjconf/201817009005}{doi:10.1051/epjconf/201817009005}
\item{} Golnik G et al. 2014 Range assessment in particle therapy based on prompt $\gamma$-ray timing measurements, \textit{Phys. Med. Biol.}  \textbf{59.18} 5399-422. 
\href{https://doi.org/10.1088/0031-9155/59/18/5399}{doi:10.1088/0031-9155/59/18/5399}
\item{} Gundacker S et al. 2014 Time resolution deterioration with increasing crystal length in a TOF-PET system, \textit{Nucl. Instrum. Methods Phys. Res. A} \textbf{737} 92-100. 
\href{https://doi.org/10.1016/j.nima.2013.11.025}{doi:10.1016/j.nima.2013.11.025}
\item{} Hueso-González F et al. 2015 First test of the prompt gamma ray timing method with heterogeneous targets at a clinical proton therapy facility, \textit{Phys. Med. Biol.} \textbf{60.16} 6247-72.
\href{https://doi.org/10.1088/0031-9155/60/16/6247}{doi:10.1088/0031-9155/60/16/6247}
\item{} Hueso-González F et al. 2018 A full-scale clinical prototype for proton range verification using prompt gamma-ray spectroscopy, \textit{Phys. Med. Biol.} \textbf{63.18} 185019.
\href{https://doi.org/10.1088/1361-6560/aad513}{doi:10.1088/1361-6560/aad513}
\item{} Kishimoto A et al. 2015 Demonstration of three-dimensional imaging based on handheld Compton camera, \textit{J. Inst.} \textbf{10.11} 11001. 
\href{https://doi.org/10.1088/1748-0221/10/11/P11001}{doi:10.1088/1748-0221/10/11/P11001}
\item{} Knopf AC and Lomax  2013 A In vivo proton range verification: a review, \textit{Phys. Med. Biol.} \textbf{58.15} 131–60.
\href{https://doi.org/10.1088/0031-9155/58/15/r131}{doi:10.1088/0031-9155/58/15/r131}
\item{} Kozlovsky B et al. 2002 Nuclear Deexcitation Gamma-Ray Lines from Accelerated Particle Interactions, \textit{Astrophys. J., Suppl. Ser.} \textbf{141.2} 523–41. 
\href{https://doi.org/10.1086/340545}{doi:10.1086/340545}
\item{} Kraan AC 2015 Range verification methods in particle therapy: underlying physics and Monte Carlo modeling, \textit{Front. Oncol.} \textbf{5} 150. 
\href{https://doi.org/10.3389/fonc.2015.00150}{doi:10.3389/fonc.2015.00150}
\item{} Krimmer J et al. 2015a Development of a Compton camera for medical applications based on silicon strip and scintillation detectors, \textit{Nucl. Instrum. Methods Phys. Res. A} \textbf{787} 98–101.
\href{https://doi.org/10.1016/j.nima.2014.11.042}{doi:10.1016/j.nima.2014.11.042}
\item{} Krimmer J et al. 2015b Collimated prompt gamma TOF measurements with multi-slit multi-detector configurations, \textit{J. Inst.} \textbf{10.01} P01011.
\href{https://doi.org/10.1088/1748-0221/10/01/P01011}{doi:10.1088/1748-0221/10/01/P01011}
\item{} Krimmer J 2017 A cost-effective monitoring technique in particle therapy via uncollimated prompt gamma peak integration, \textit{Appl. Phys. Lett.} \textbf{110.15} 154102. 
\href{https://doi.org/10.1063/1.4980103}{doi:10.1063/1.4980103}
\item{} Krimmer J et al. 2018 Prompt-gamma monitoring in hadrontherapy: A review, \textit{Nucl. Instrum. Methods Phys. Res. A} \textbf{878} 58-73. 
\href{https://doi.org/10.1016/j.nima.2017.07.063}{doi:10.1016/j.nima.2017.07.063}
\item{} Marcatili S et al. 2019 A 100 ps TOF Detection System for On-Line Range-Monitoring in Hadrontherapy, \textit{IEEE Nuclear Science Symposium and Medical Imaging Conference} 1–4.
\href{https://doi.org/10.1109/NSS/MIC42101.2019.9059815}{doi:10.1109/NSS/MIC42101.2019.9059815}
\item{} Marcatili S et al. 2020 Ultra-fast prompt gamma detection in single proton counting regime for range monitoring in particle therapy, accepted for publication on \textit{Phys. Med. Biol.}.
\href{https://doi.org/10.1088/1361-6560/ab7a6c}{doi:10.1088/1361-6560/ab7a6c}
\item{} Muñoz E et al. 2017 Performance evaluation of MACACO: a multilayer Compton camera, \textit{Phys. Med. Biol.} \textbf{62.18} 7321-41.
\href{https://doi.org/10.1088/1361-6560/aa8070}{doi:10.1088/1361-6560/aa8070}
\item{} Paganetti H 2012 Range uncertainties in proton therapy and the role of Monte-Carlo simulations, \textit{Phys. Med. Biol.} \textbf{57.11} 99-117. 
\href{https://doi.org/10.1088/0031-9155/57/11/R99}{doi:10.1088/0031-9155/57/11/R99}
\item{} Parodi K et al. 2018 In vivo range verification in particle therapy, \textit{Med. Phys.} \textbf{45.18} 1036-50.
\href{https://doi.org/10.1002/mp.12960}{doi:10.1002/mp.12960}
\item{} Pausch G et al. 2016 Scintillator-Based High-Throughput Fast Timing Spectroscopy for Real-Time Range Verification in Particle Therapy, \textit{IEEE Trans. Nucl. Sci.} \textbf{63.2} 664–672. 
\href{https://doi.org/10.1109/TNS.2016.2527822}{doi:10.1109/TNS.2016.2527822}
\item{} Pausch G et al. 2020 Detection systems for range monitoring in proton therapy: Needs and challenges, \textit{Nucl. Instrum. Methods Phys. Res. A} \textbf{954} 161227. 
\href{https://doi.org/10.1016/j.nima.2018.09.062}{doi:10.1016/j.nima.2018.09.062}
\item{} Perali I et al. 2014 Prompt gamma imaging of proton pencil beams at clinical dose rate, \textit{Phys. Med. Biol.} \textbf{59.19} 5849–71.
\href{https://doi.org/10.1088/0031-9155/59/19/5849}{doi:10.1088/0031-9155/59/19/5849}
\item{} Petzoldt J et al. 2016 Characterization of the microbunch time structure of proton pencil beams at a clinical treatment facility, \textit{Phys. Med. Biol.} \textbf{61.6} 2432-56. 
\href{https://doi.org/10.1088/0031-9155/61/6/2432}{doi:10.1088/0031-9155/61/6/2432}
\item{} Pinto M et al. 2014 Design optimisation of a TOF-based collimated camera prototype for online hadrontherapy monitoring, \textit{Phys. Med. Biol.} \textbf{59.24} 7653-74.
\href{https://doi.org/10.1088/0031-9155/59/24/7653}{doi:10.1088/0031-9155/59/24/7653}
\item{} Pinto M et al. 2016 Assessment of Geant4 Prompt-Gamma Emission Yields in the Context of Proton Therapy Monitoring, \textit{Front. Oncol. } \textbf{6}. 
\href{https://doi.org/10.3389/fonc.2016.00010}{doi:10.3389/fonc.2016.00010}
\item{} Priegnitz M et al. 2015 Measurement  of  prompt  gamma  profiles  in inhomogeneous targets with a knife-edge slit camera during proton irradiation, \textit{Phys. Med. Biol.} \textbf{60.12} 4849-71. 
\href{https://doi.org/10.1088/0031-9155/60/12/4849}{doi:10.1088/0031-9155/60/12/4849}
\item{} Richter C et al. 2016 First clinical application of a prompt gamma based in vivo proton range verification system, \textit{Radiother.  Oncol.} \textbf{118} 232–37.
\href{https://doi.org/10.1016/j.radonc.2016.01.004}{doi:10.1016/j.radonc.2016.01.004}
\item{} Roellinghoff F et al. 2011 Design of a Compton camera for 3D prompt-imaging during ion beam therapy, \textit{Nucl. Instrum. Methods Phys. Res. A}  \textbf{648} 20–23.
\href{https://doi.org/10.1016/j.nima.2011.01.069}{doi:10.1016/j.nima.2011.01.069}
\item{} Smeets J et al. 2012 Prompt gamma imaging with a slit camera for real-time range control in proton therapy, \textit{Phys. Med. Biol.} \textbf{57.11} 3371–405. 
\href{https://doi.org/10.1088/0031-9155/57/11/3371}{doi:10.1088/0031-9155/57/11/3371}
\item Thirolf PG et al. 2017 A Compton camera prototype for prompt gamma medical imaging, \textit{EPJ Web Conf.} \textbf{117} 05005. 
\href{https://doi.org/10.1051/epjconf/201611705005}{doi:10.1051/epjconf/201611705005}
\item{} Verburg JM et al. 2012 Simulation of prompt gamma-ray emission during proton radiotherapy, \textit{Phys. Med. Biol.} \textbf{57.17} 5459–72. 
\href{https://doi.org/10.1088/0031-9155/57/17/5459}{doi:10.1088/0031-9155/57/17/5459}
\item{} Verburg JM and Seco J 2014 Proton range verification through prompt gamma-ray spectroscopy, \textit{Phys. Med. Biol.} \textbf{59.23} 7089–106. 
\href{https://doi.org/10.1088/0031-9155/59/23/7089}{doi:10.1088/0031-9155/59/23/7089}
\item{} Werner T et al., 2019 Processing of prompt gamma-ray timing data for proton range measurements at a clinical beam delivery, \textit{Phys. Med. Biol.} \textbf{64} 105023-43.
\href{https://doi.org/10.1088/1361-6560/ab176d}{doi:10.1088/1361-6560/ab176d}
\item{} Xie Y et al. 2017 Prompt Gamma Imaging for In Vivo Range Verification of Pencil Beam Scanning Proton Therapy, \textit{Int. J. Radiat. Oncol. Biol. Phys.} \textbf{99.1} 210–218. 
\href{https://doi.org/10.1016/j.ijrobp.2017.04.027}{doi:10.1016/j.ijrobp.2017.04.027}
\end{harvard}
\end{document}